\documentstyle[prl,floats,aps,twocolumn,epsf,graphicx]{revtex}
\begin{document}
\twocolumn[\hsize\textwidth\columnwidth\hsize\csname
@twocolumnfalse\endcsname

%\begin{document}
\title{Survival probabilities in time-dependent random walks}
\author{Ehud Nakar and Shahar Hod}
\address{The Racah Institute of Physics, The Hebrew University, Jerusalem 91904, Israel}
\date{\today}
\maketitle

\begin{abstract}

\ \ \ We analyze the dynamics of random walks in which the jumping probabilities 
are periodic {\it time-dependent} functions. In particular, we determine the 
survival probability of biased walkers who are drifted towards an absorbing boundary. 
The typical life-time of the walkers is found to decrease with an increment of the 
oscillation amplitude of the jumping probabilities. 
We discuss the applicability of the results in the context of complex adaptive 
systems.
\end{abstract}
\bigskip

]

Random walk is one of the most ubiquitous concepts of statistical
physics. In fact, it finds applications in virtually every area of
physics (see e.g.,\cite{BaNi,Kam,FeFrSo,Wei,AvHa,DiDa} and
references therein). Random walks in the presence of absorbing
traps are much studied in recent years as models for
absorbing-state phase transitions \cite{MaDi,Hin}, polymer
adsorption \cite{BeLo}, granular segregation \cite{FaFu}, and the
spreading of an epidemic \cite{GrChRo}.

One of the main characteristics of a random walk in the presence of 
an absorbing boundary is the survival probability $S(t_0)$ of the the 
walkers, the probability that a walker has not reached the 
absorption point before $t_0$. 
When the walkers are drifted towards the absorbing point, the survival 
probability falls exponentially at asymptotically late-times 
$S(t)\propto exp(-t/\tau_s)$, where the characteristic life-time $\tau_s$ 
depends on the drift velocity (see e.g. \cite{BaGoLu}).

The problem of random walks with an absorption boundary  has recently been 
extended to include situation in which the jumping probabilities of the 
walkers are {\it time-dependent} \cite{Hod03}. The solution 
presented in \cite{Hod03} accounts for the binary (alternating) case, 
namely $p(t)={1 \over 2}-\epsilon_p +(-1)^t A$, where $p(t)$ is the 
probability to step to the right at time $t$. Here $A$ is the oscillations 
amplitude, and $\epsilon_p>0$ represents a net drift towards the absorbing boundary (the 
period of such binary oscillations is $T=2$). 
It was found \cite{Hod03} that the characteristic life-time of the walkers $\tau_s$ 
depends on the amplitude $A$, in a non-trivial way. 
In particular, $\tau_s$ was shown to decrease monotonically with 
the increment of $A$. While this analysis provides a useful 
insight into the behavior of such time-dependent random walks, it is 
of one's interest to explore the general case time-dependent 
jumping probabilities with $T \neq 2$. This is the main goal of the 
present work.

In addition to the intrinsic interest in such time-dependent random walks, our 
work may find direct applications in many complex physical, biological, and 
economical systems. In fact,  the main motivation for the introduction of 
time-dependent random walks in \cite{Hod03} was its applicability 
in the flourishing field of complex adaptive systems. 
In the well-studied model of the  minority game (MG) \cite{ChaZha}, 
and its evolutionary version (EMG) \cite{John1} 
(see also\cite{DhRo,BurCev,LoHuJo,HuLoJo,HaJeJoHu,BuCePe,LoLiHuJo,LiVaSa,SaMaRi,HodNak1,Coo,HodNak2,CaCe}), 
it was found that the winning probabilities of the agents display 
a periodic behavior in time. This implies that the survival probabilities of the 
agents are well-described by a model of a periodic time-dependent random walk with an 
absorbing boundary \cite{Hod03}. The analytical model presented in \cite{Hod03} (with $T=2$) 
provides an elegant explanation for the intriguing phase-transition 
observed in the EMG \cite{HodNak1} (from self-segregation to clustering as the prize-to-fine ratio drops 
below some critical value). 
However, numerical studies of the EMG \cite{HodNak2} have indicated that 
the period of the oscillations (in the winning probabilities of the agents) depends on the 
specific parameters of the system. Extending the analysis of \cite{Hod03} to the general case of 
time-dependent jumping probabilities with $T \neq 2$ is therefore 
highly motivated.

We consider a time dependent random walker who is drifted towards an absorption boundary, 
located at $-d$ (where $d>0$). The drift towards 
the absorption point may be attributed to two distinct reasons: (i) A smaller (average) 
probability to take a step to the right (away from the
absorption point). That is, $\langle p(t) \rangle=1/2-\epsilon_p$ where $\epsilon_p >0$, 
and (ii) a smaller step size $r$ to the right, $0\leq r<1$ 
(where the step size to the left is scaled to $1$). This kind of drift is 
characterized by the parameter $\epsilon_r \equiv 1/(1+r)-{1 \over 2}$. 
In the unbiased case ($\epsilon_p=\epsilon_r=0$) 
there is no net drift, and the survival probability is well 
known to scale as an inverse power-law: $S(t)\propto t^{-1/2}$. In the
presence of a drift ($\epsilon_p >0$ and/or $\epsilon_r>0$), and with a 
constant (time-{\it independent}) jumping probabilities, one finds 
(see e.g. \cite{BaGoLu})

\begin{equation}
\label{EQ S_t}
S(t)\propto t^{-{3 \over 2}} e^{-t/{\tau }_{s}}\  .
\end{equation}
This result holds true in the {\it binary} time-dependent model \cite{Hod03} as well, 
in which case the jumping probability is given by the alternating 
function $p(t)={1 \over 2}-\epsilon_p + (-1)^tA$.

Below we explore the time-dependence of the survival probability $S(t)$ for a general 
periodic $p(t)$. Let

\begin{equation}
\label{eq r(t)}
p(t)={1 \over 2}+F(t)\  ,
\end{equation}
where $F(t)$ is a periodic function, with a period $T$. 
The mean of $F$ (over one period) is given by 

\begin{equation}
\label{EQ F}
\langle F \rangle \equiv (1/T)\sum ^{T}_{i=1}F(i)=-\epsilon_p\  .
\end{equation} 

In the time-independent case it has been established 
that the probability $B(t)$ for a random walker 
(with {\it no} absorbing boundary) to be located on the 
right hand side [$x(t)>0$] at the time $t$ is given by $B(t) \propto exp[-t/\tau_B]$, with 
$\tau _B=\tau_s$ (see e.g., \cite{BaGoLu}). 
Motivated by the equality $\tau _B=\tau_s$ in the time independent case, 
we will find the asymptotic form of $\tau_B$ in the general case of 
time-dependent jumping probabilities. We will establish 
the relation $\tau_B(T=2)\approx \tau_s(T=2)$ analytically, and demonstrate numerically 
that $\tau_B(T \neq 2)\approx \tau_s(T \neq2)$.

We first consider the case in which the period $T$ of the oscillations is a natural
number. Let $\omega(t)$ be the number of right steps
taken by the walker out of $t$ steps. The walker is located at the 
right hand side [$x(t)>0$] if $\omega>\omega_c$, where  

\begin{equation}
\label{EQ w_{c}} 
\omega_c={{t-d} \over {1+r}}\simeq t({1 \over 2}+\epsilon_r)\  ,
\end{equation}
in the  $t\gg d$, $T$ limit.

Note that $\omega$ is a sum of $T$ independent binomial distributions with
a success probability $p_{i}=p(t=i)$. After time $t\gg T$ a step is chosen
$\simeq t/T$ times from each one of these distributions. Thus, for
$t\gg T$, $\omega$ is distributed normally with an average of 
$\mu_w(t)=t({1 \over 2}-\epsilon_p)$, and a variance 
$\sigma^2_{\omega}={t \over T}\sum^{T}_{i=1}p_{i}(1-p_{i})=t(1-4 \langle F^2 \rangle)/4$, where 
 
\begin{equation}
\label{EQ F2}
\langle F^2 \rangle \equiv (1/T)\sum ^{T}_{i=1}F^2(i)\  .
\end{equation}
Hence, $B(t)$ can be approximated by this
normal distribution, with the condition $\omega(t)>\omega_c(t)$\
 
\begin{eqnarray}
\label{EQ B(t)}
B(t)=\frac{1}{\sqrt{2\pi \sigma ^{2}}}\int ^{\infty
}_{w_{c}}exp\left[ -\frac{(x-\mu )^{2}}{2\sigma ^{2}}\right]
dx\nonumber \\
=\frac{1}{\sqrt{\pi }}\int ^{\infty }_{\sqrt{t/\tau
_{B}}}e^{-x^{2}}dx=\frac{1}{2}erfc(\sqrt{t/\tau _{B}})\  ,
\end{eqnarray}
where

\begin{equation}
\label{EQ tauB}
\tau _B=\frac{2\sigma ^2t}{(w_c-\mu)^{2}}=\frac{1-4\langle F^2\rangle}{2(\epsilon_p+\epsilon_r)^2}\  ,
\end{equation}
and \( erfc \) is the complementary error function. The complementary error function 
can be approximated as $erfc(\sqrt{t/\tau _{B}})\simeq e^{-t/\tau _{B}}/\sqrt{\pi t/\tau_{B}}$ at 
asymptotically large times. One therefore finds

\begin{equation}
B(t)\propto e^{-t/\tau_{B}}\  .
\end{equation}

The model of binary time-dependent jumping probabilities [$F(t)=-\epsilon_p+A(-1)^t$] 
was solved {\it exactly} in \cite{Hod03}, the expression for $\tau_s$ is given by 
Eq. ($16$) of \cite{Hod03}. Expanding this expression for small $\epsilon_p$ and 
$\epsilon_r$, one finds

\begin{equation}
\label{EQ tauB_T2}
\tau_s(T=2)\simeq {{1-4A^2} \over {2(\epsilon_p+\epsilon_r)^2}}\  .
\end{equation}
This agrees with Eq. (\ref{EQ tauB}) in the limit $\epsilon_p \ll 1$ (or equivalently, 
$\tau_s \gg 1$). 
One therefore finds $\tau_s(T=2) \simeq \tau_B(T=2)$ in the $\tau_s \gg 1$ limit.

The generalization of the above analysis for an arbitrary period $T$ is 
straightforward. If $T=m/n$ is a rational number, then $F(t)$ could be replaced by an
equivalent function with a period $m$ which is a natural number. 
If $T$ is an irrational, then the all range of
values of $F(0<t<T)$ is sampled (over many periods of the oscillations). 
In this case, one should replace Eqs. (\ref{EQ F}) and (\ref{EQ F2}) by

\begin{equation}
\langle F \rangle=\frac{1}{T}\int_0^T F(t)dt=-\epsilon\  ,
\end{equation}
and

\begin{equation}
\langle F^2 \rangle=\frac{1}{T}\int_0^T F^2(t)dt\  .
\end{equation}

\begin{figure}[tbh]
\centerline{\epsfxsize=9cm \epsfbox{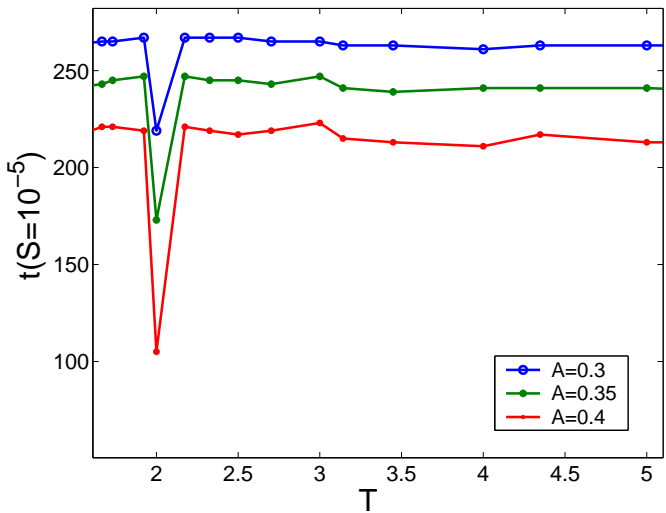}} 
\caption{The step number for which the survival probability has fallen to $S=10^{-5}$, 
as a function of the oscillation period $T$, and for 
three different values of $A$. The jumping probabilities $p(t)$ are characterize by harmonic 
oscillations, and are given by Eq. (\ref{EQ r_sinus}). 
We use $N=10^8$ walkers, $\epsilon_p=0.1$, $r=1$, and
$d=2$. $S(t)$ is independent of the oscillation period $T$, except in the 
unique case of binary oscillations with $T=2$ [in which case $S(t)$ is 
smaller, in agreement with the analytical prediction.] 
In addition, $S(t)$ decreases monotonically with the increment of the oscillations amplitude $A$.}  
\label{Fig f1}
\end{figure}

In order to confirm the analytical predictions (in the $T \neq 2$ case), 
we perform numerical simulations of (discrete) random walks. 
We first consider the case of an harmonic function $p(t)$ of the form

\begin{equation}
\label{EQ r_sinus}
p(t)={1 \over 2}-\epsilon_p -Acos(\frac{2\pi t}{T})\  ;\   T>1\  .
\end{equation}
In this case $\langle F \rangle =-\epsilon_p$, and

\begin{equation}
\label{EQ F2_sinus}
\langle F^2 \rangle=\frac{1}{4}\left\{
\begin{array}{c}
\epsilon^2_p+A^{2}\  ;\quad T=2\\
\epsilon^2_p+(A/\sqrt{2})^{2}\  ;\quad T\neq 2
\end{array}\right.
\end{equation}
This implies that for $T \neq 2$, $\tau _{B}$ is {\it independent} of the period $T$ of the 
oscillations. Moreover, we find that 
$\tau_B(T \neq 2,A/\sqrt{2})=\tau _B(T=2,A) $. As we know that our
approximation holds true for $T=2$ (The exact 
shape of $F$ is irrelevant in this case), one should compare 
the survival probabilities in the $T=2$ and $T \neq 2$ cases. 
Figure \ref{Fig f1} depicts $t(S=10^{-5})$, the step number in which the survival 
probability has fallen to $S=10^{-5}$, as a function of the oscillation period $T$ 
(and for three different values of the oscillation amplitude $A$). 
As predicted by Eq. (\ref{EQ F2_sinus}), the survival probabilities are almost 
independent of the period $T$, except in the unique case of binary oscillations with $T=2$. 
Moreover, $S(t;T=2) < S(t;T \neq 2)$, in agreement with Eqs. (\ref{EQ tauB}) and (\ref{EQ F2_sinus})]. 
In addition, the numerical simulations indicate that the survival probability decreases monotonically with an increment 
in the oscillation amplitude $A$, in agreement with the analytical prediction 
Eq. (\ref{EQ tauB}).

\begin{figure}[tbh]
\centerline{\epsfxsize=9cm \epsfbox{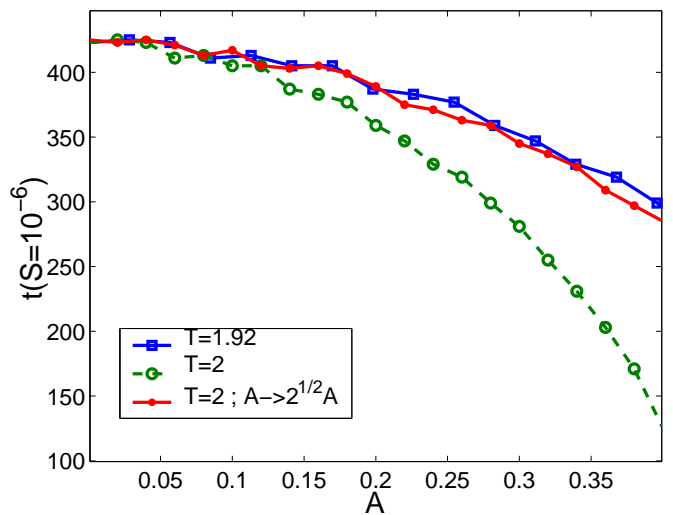}} 
\caption{The step number in which $S=10^{-6}$ as a function of the oscillations 
amplitude $A$, and for two different values of the period $T$. The results confirm 
the analytical prediction according to which $S(t,T=2,A)=S(t,T\neq 2,\sqrt{2}A)$, where 
$p(t)$ is an harmonic function. The parameters used are the same as in Fig. 1.}
\label{Fig f2}
\end{figure}

In Figure \ref{Fig f2} we confirm the relation $S(t,T\neq 2,A)=S(t,T= 2,\sqrt{2}A)$ in the 
harmonic case. The figure displays the step number $t(S=10^{-6})$ for which the survival probabilities 
have fallen to $10^{-6}$. It is clear that $t(S=10^{-6})$
for $T=2$ agrees with the corresponding value in the $T \neq 2$ case, provided one 
uses the transformation $A\rightarrow \sqrt{2}A$.

\begin{figure}[tbh]
\centerline{\epsfxsize=9cm \epsfbox{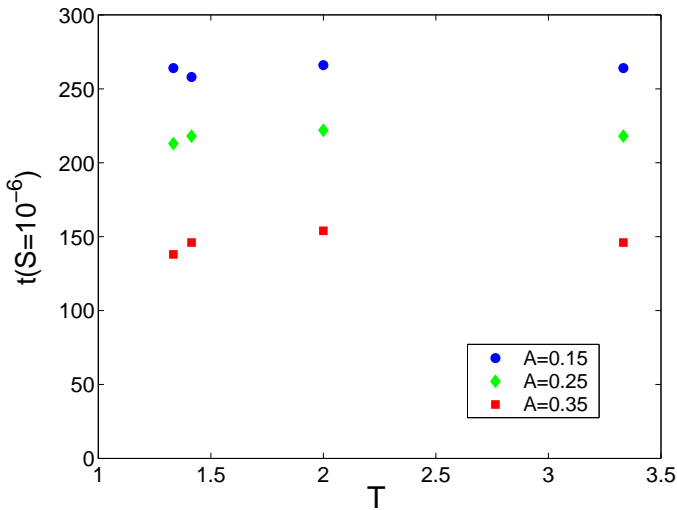}} 
\caption{The step number for which the survival probability has fallen to $S=10^{-6}$, 
as a function of the oscillation period $T$, and for 
three different values of $A$. $F(t)$ is a square wave with amplitude $A$, 
and a mean value $-\epsilon_p$. 
We use $N=10^9$ walkers, $\epsilon_p=0.1$, $R=0.9$, and
$d=2$. We find that in this case, $S(t)$ is independent of the oscillation period $T$, 
in agreement with the analytical results.}
\label{Fig f3}
\end{figure}

We further check the validity of the analytical results for $F(t)$ in the form of 
a square wave of amplitude $A$, and a mean value $-\epsilon_p$. In this case one finds 
$\langle F^2 \rangle=\epsilon^2_p+A^2$ for any $T$. This implies that $\tau_B(T)$ should be 
{\it independent} of the period $T$ of the oscilations (and in particular, 
$\tau_s(T) =\tau_s(T=2)$, where $\tau_s(T=2)$ was derived analytically in \cite{Hod03}). 
Figure \ref{Fig f3} depicts $t(S=10^{-6})$, the step number for which the survival probability 
has fallen to $S=10^{-6}$, as a function of the period $T$. We display results for three 
different values of the oscillation amplitude $A$. The numerical results confirm the finding according to 
which $S(t;T)$ is independent of the value of $T$. 
Furthermore, the survival probability $S(t;T)$ decreases with increasing amplitude $A$, in 
agrrement with Eq. (\ref{EQ tauB}).

In summary, in this paper we studied the survival probabilities of biased (drifted) 
random walkers, whose jumping probabilities are {\it time-dependent}. This is 
an extension of the binary case (with $T=2$), studied in \cite{Hod03}. 
The long-time asymptotic survival probability is dominated by an exponential 
fall $S(t)\propto exp[-t/\tau_s]$. 
We have found a simple approximation for $\tau_s$ 
[see Eq. (\ref{EQ tauB})], which solely depends on the first and the second moments 
of the oscillations ($\langle F \rangle$ and $\langle F^2 \rangle$), and on the step-size drift
parameter $\epsilon_r$. 

Our analytical results imply that the characteristic life-time of the 
walkers $\tau_s$ {\it decreases} with increasing $\langle F^2 \rangle$. 
The larger are the temporal-oscillations in the jumping probabilities $p(t)$, the 
smaller is the survival probability. This result generalizes the one derived 
in \cite{Hod03} for the binary case (with $T=2$), in which case $\langle F^2 \rangle=A^2+{\epsilon^2_p}$.

The qualitative nature of the anti-correlation between the survival probability $S(t)$ and 
the oscillation amplitude $\langle F^2 \rangle$ is rather simple. 
It is a direct consequence of the positive correlation between $\sigma_{\omega}(t)$ 
(the dispersion in the access of right-steps taken out of $t$ steps) and $S(t)$ 
in the presence of a negative drift (towards the absorption boundary). 
In the presence of a net drift, $\mu_{\omega}$ (the average value of $\omega$) 
is smaller than $\omega_c$. A large $\sigma_{\omega}(t)$ is therefore required 
in order to survive, that is in order to have $\omega > \omega_c$ at asymptotically late-times \cite{Note}. 

Finally, the present analysis provides a direct explanation for the underlying mechanism responsible for 
the intriguing phase-transition observed in the evolutionary
minority  game \cite{HodNak1,John1}. In this model, the winning probabilities of the 
agents $p(t)$ where shown to display temporal oscillations \cite{HodNak2}. 
The amplitude and period of these oscillations depend 
on the various parameters of the model (such as the price-to-fine ratio). 
In \cite{Hod03} a toy-model of time-dependent jumping probabilities with a period of $T=2$ was 
used to reveal the physical source for the phase-transition. In the present work we 
have shown that the anti-correlation between the characteristic life-time $\tau_s$ and the 
oscillations amplitude $\langle F^2 \rangle$ is a generic feature, {\it independent} of the period $T$ 
of the oscillations. 
Thus, our analysis (for generic values of $T$) lends support for the general applicability 
of the conclusions presented in \cite{Hod03}.

\bigskip
\noindent
{\bf ACKNOWLEDGMENTS}
\bigskip

The research of SH was supported by G.I.F. Foundation. The research of EN was supported by the 
Horowitz foundation, and through the generosity of the Dan David Prize Scholarship 2003. We would also 
like to thank Yonatan Oren.


\begin{thebibliography}{99}
\bibitem{Note} We note that in the unbiased case (with no drift), one finds $\mu_{\omega} > \omega_c$.
A small $\sigma_{\omega}(t)$ therefore implies large survival probability. Thus, 
$\langle F^2 \rangle$ is positively correlated with $S(t)$ in the unbiased case, in agreement with 
the results of \cite{Hod03} for the binary case.
\end{thebibliography}
\end{document}